\documentclass[12pt]{article}
\usepackage{amssymb,amsmath,epsfig,graphicx}
\usepackage{amsfonts}
\usepackage{amsthm}
\usepackage{caption}
\usepackage{epstopdf}
\begin{document}
\title{\bf Stellar Hydrostatic Equilibrium Compact Structures in $f(\mathcal{G},T)$ Gravity}
\author{M. Farasat Shamir \thanks{farasat.shamir@nu.edu.pk}
and Mushtaq Ahmad \thanks{mushtaq.sial@nu.edu.pk}\\\\ National University of Computer and
Emerging Sciences,\\ Lahore Campus, Pakistan.}

\date{}

\maketitle
\begin{abstract}
In this paper, stellar hydrostatic equilibrium configuration of the compact stars (neutron stars and strange stars) has been studied for $f(\mathcal{G},T)$ gravity model, with $\mathcal{G}$ and $T$ being the Gauss-Bonnet invariant and the trace of energy momentum tensor, respectively. After having derived the hydrostatic equilibrium equations for $f(\mathcal{G},T)$ gravity, the fluid pressure for the neutron stars and the strange stars has been computed by implying two equation of state models corresponding to two different existing compact stars. For the $f(\mathcal{G},T)=\alpha{\mathcal{G}^n+\lambda{T}}$ gravity model, with $\alpha$, $n$, and $\lambda$ being some specific constants, substantial change in the behavior of the physical attributes of the compact stars like the energy density, pressure, stellar mass, and total radius has been noted with the corresponding change in $\lambda$ values. Meanwhile, it has been shown that for some fixed central energy density and with increasing values of $\lambda$, the stellar mass both for the neutron stars and the strange stars increases, while the total stellar radius $R$ exhibits the opposite behavior for both of the compact stars. It is concluded that for this $f(\mathcal{G},T)$ stellar model, the maximum stellar mass can be boosted above the observational limits.

\end{abstract}


\section{Introduction}
The creation of the compact stars originates from the conclusive phase of the gravitational collapse occurring during the evolution process of massive ordinary stars due to their failure to endure their stability as they would have, at that time, finished their nuclear fuel essential for their existence. This is the moment when the internal radial pressure no longer is adequate to overcome the gravitational force and this makes the star with no other option but to collapse and leaving behind compact objects like the neutron stars, strange stars, white dwarfs, or a black hole, as the leftovers. These compact stars are made of some stiff matter with high densities responsible for their massive structures. This is what which makes these compact stars as neutron stars or strange stars, being usually highly massive objects with very small radii. These structures are geometrically described by the Tolman–Oppenheimer–Volkoff equations (TOV) \cite{OV}, the solution of which becomes too important to study the complexities involved in these massive small-sized structures. These equations systematically combine the mass, radius, energy density, and pressure and show how the pressure and the energy density variate with respect to the mass and radius of the compact star. In case of compact stars (in particular the neutron stars), the internal pressure equivalent to the gravitational pressure is in fact the pressure being exhibited from some degeneracy of fermions.\par
 In general relativity (GR), the comprehensive investigations of the hydrostatic equilibrium structures become significant
 for the reason of compact stars containing large mass and density.

 This may begin with the general relativistic approach with the assumption of TOV equations for the non-rotating spherically symmetric hydrostatic equilibrium, as follows

\begin{equation}\label{OV}
\frac{dp}{dr}=-G\frac{(\rho{c^2}+p)(mc^2+4p\pi{r^3})}{r^2c^4-2Gmc^2r}~~~~\text{and}~~~~\frac{dm}{dr}=4\rho{\pi}r^2,
\end{equation}
where $p$, $\rho$, and $m$ is the radial pressure, energy density, and the stellar mass of the star respectively, being dependent on radial coordinate $r$.
At the boundary $r=R$ of the compact star, the total stellar mass of the compact stars is calculated as 
\begin{equation}\label{OVMass}
M(R)=\int_{0}^{R}4\pi{r^2}\rho{d}{r}.
\end{equation}
For the purpose to close the above coupled equations, an appropriate choice of the equation of state (EoS) becomes significant here, which eventually exhibits the proportional relation between the energy density and the pressure.
To investigate the hydrostatic stellar equilibrium, combination of these TOV equations with the EoS when
solved with some suitable numerical methods, results into distinguished stellar configurations of the
compact stars depending on divisions of low and high densities \cite{Negele, AAkmal}.

Schwarzschild \cite{Schwild} presented the very first spherically symmetric exact solution
to the Einstein field equations, and as an outcome of the solution was the spacetime
singularity which further emerged with the idea of the black hole. The second exact non-trivial
solution explored the bounds on the compactness parameter i.e.,
$\mu{(R)}=\frac{2M(R)}{R}<\frac{8}{9}$ in stellar hydrostatic equilibrium
for  static, spherically symmetric compact configurations
\cite{Buchdahl}. Exploring compacts stars, specifically the neutron stars,and the strange stars, has turned to be the hot pursuit in Astrophysics. Baade and Zwicky \cite{Baade} investigated stellar
structures and explored that supernova may convert itself to compact dense object which came true later at the time of discovery of pulsars (which are highly magnetized revolving neutrons) \cite{Longair, Ghosh}.
Mak and Harko \cite{MaK} presented standard models for the spherically symmetric
compact structures and found exact
solutions. These solutions were helpful in determining the physical attributes like the energy
density, tangential and the radial pressures. They concluded that inside
these compact stars, these physical parameters  remained finite and positive.
Hossein et al. \cite{Hossein} explored the role of the cosmological constant on anisotropic
stars. Different distinguishing physical aspects such as the stellar mass, radius, and moment of inertia related to the neutron stars have been investigated and a comprehensive comparative study of GR with modified theories of gravity \cite{32d} has been presented. Some interesting related studies on the compact configurations of the neutron stars, rotating slowly in $R^2$ gravity are accomplished by using distinct hadronic and a the strange matter EOS parameter \cite{32e}. One can find some fascinating related research works in \cite{NJO7}-\cite{Capoz2}.

Modified theories of gravity have equipped the researchers with different astronomical techniques to study the reasons behind so called accelerating expansion of this universe. Harko and collaborators \cite{HarkoA} were the first who came up with the idea of implicit-explicit couplings of  curvature and matter by poising a new modified theory of gravity, known as the $f(R,T)$ gravity.
Some interesting works regarding modified theories of gravity may be noted
in \cite{NJO}-\cite{Sta} and in \cite{Sir&Me3,Sir&SZ}.
In the recent years, researchers have presented some generalized approaches in modified
Gauss-Bonnet gravity. In research work, Sharif and Ikram \cite{sharif.ayesha}
presented yet another modified $f(\mathcal{G},T)$ theory of gravity and
worked out different bounds on energy for the Friedmann-Robertson-Walker
(FRW)metric.  They concluded that some specific test objects follow
non-geodesic geometrical aspects when presented with some extra dynamics.
 We used the Noether symmetry approach to present the exact solutions in $f(\mathcal{G},T)$
 gravity \cite{Sir&M}. Also, applying the same approach, we detemined some cosmological sustainable $f(\mathcal{G},T)$
gravity models to anisotropic background for locally rotationally symmetric
Bianchi type $I$ universe \cite{Sir&Me2}. We inferred that without involving cosmological constant, some
Gauss-Bonnet dependent particular models may be used for the
reconstruction of $\Lambda$CDM cosmology.\par
Momeni and Myrzakulov in \cite{momeniD} constructed a neutron star model based on a stringy inspired Gauss-Bonnet modification of classical gravity and derived the modified forms of the TOV equations for $f(\mathcal{G})$ gravity and they concluded that the dynamics of the metric functions were modified due to the Gauss-Bonne term effects while the static equations remained without any change. Artyom  and Collabarators in \cite{AsCapOd}  considered Quark star models with realistic equation of state in non-perturbative $f(R)$ gravity and obtained the mass-radius relation for $f(R)=R+R^2$. Further, they explored that it was possible to differentiate the modified theories of gravity from GR due to the existence of gravitational redshift of the thermal spectrum emerging from the surface of the star. Cemsinan Deliduman et al. \cite{Cemsinan} studied the structure of neutron stars in $R+\beta{R^{\mu\nu}R_{\mu\nu}}$ gravity with perturbative approach and obtained the mass-radius relations for the six representative EoS parameters and found  subsequently different results as compared to GR.\par
Moraes et al. \cite{MOR} investigated the hydrostatic equilibrium stellar configuration of neutron stars
and strange stars in $f(R,T)$ theory of gravity and computed their corresponding fluid pressures from the EoS $p=\omega{\rho^{5/3}}$ and $p=0.28(\rho-4\beta)$ by starting with the derivation of TOV equations for $f(R,T)$ theory of gravity. They used the functional form of $f(R,T)=R+2\lambda{T}$ and discussed the corresponding change in the energy density, radial pressure, and stellar mass against different values of $\lambda$. Moreover, they concluded that the EoS cannot be eradicated provided the maximum stellar mass within GR holds under the observed pulsars limits.\par

The setup of this paper is administered as follows: In Section $2$, a brief description of the modified $f(\mathcal{G}, T)$ theory of gravity with its fundamental formalism  has been provided. Section $3$, presents the equations of stellar configurations, the boundary constraints, and the EoS  for the analysis of the compact stars (neutron stars, strange stars) in $f(\mathcal{G}, T)$ gravity.  Section $4$, includes the presentation of equilibrium configurations of the neutron stars and the strange stars in $f(\mathcal{G}, T)$ gravity. Section $5$, contains the conclusive remarks with brief discussions on the results.

\section{The $f(\mathcal{G},T)$ Gravity}
The proposed modified Gauss-Bonnet $f(\mathcal{G},T)$ gravity \cite{sharif.ayesha} makes use of gravitational aspect of the action depending upon a generic function of $\mathcal{G}$, the Gauss-Bonnet invariant, defined as
\begin{equation}\label{GBT}
  \mathcal{G}=R^2-4R_{\zeta\eta}R^{\zeta\eta}+R_{\zeta\eta\gamma\delta}R^{\zeta\eta\gamma\delta},
\end{equation}
and $T$, the trace of energy momentum tensor $T_{\alpha\beta}$. By taking into account the matter Lagrangian energy density $\mathcal{L}_{M}$, one can read the total action as
 \begin{equation}\label{action}
\mathcal{Z}= \frac{1}{2{\kappa}^{2}}\int d^{4}x
\sqrt{-g}[R+f(\mathcal{G},\mathrm{\textit{T}})]+\int
d^{4}x\sqrt{-g}\mathcal{L}_{M},
\end{equation}
where $g$ assumes the metric determinant,
$\kappa$ being the coupling constant, and $R$ being the expression for the Ricci Scalar.
 Variating Eq.(\ref{action}) with
respect to $g_{\xi\eta}$, the metric tensor, the modified field equations read
\begin{eqnarray}\nonumber
G_{\xi\eta}&=&\mathrm{\textit{T}}_{\xi\eta}+[2Rg_{\xi\eta}\nabla^{2}+
2R\nabla_{\xi}\nabla_{\eta}+4g_{\xi\eta}R^{\mu\nu}\nabla_{\mu}\nabla_{\nu}+
4R_{\xi\eta}\nabla^{2}-\\\nonumber
&&4R^{\mu}_{\xi}\nabla_{\eta}\nabla_{\mu}-4R^{\mu}_{\eta}\nabla_{\xi}\nabla_{\mu}-4R_{\xi\mu\eta\nu}\nabla^{\mu}\nabla^{\nu}]f_{\mathcal{G}}+
\frac{1}{2}g_{\xi\eta}f-[\mathrm{\textit{T}}_{\xi\eta}+\Theta_{\xi\eta}]\times\\
&&f_{\mathrm{\textit{T}}}-[2RR_{\xi\eta}-4R^{\mu}_{\xi}R_{\mu\eta}-4R_{\xi\mu\eta\nu}R^{\mu\nu}+2R^{\mu\nu\delta}_{\xi}R_{\eta\mu\nu\delta}]
f_{\mathcal{G}},\label{4_eqn}
\end{eqnarray}
where the box $\Box=\nabla^{2}=\nabla_{\xi}\nabla^{\xi}$ expresses the
d'Alembertian operator with $\nabla_{\xi}$ as the covariant derivative taken along the symmetric connection related to $g_{\xi\eta}$ , ${G}_{\xi\eta}=R_{\xi\eta}-\frac{1}{2}g_{\xi\eta}R$ shows
Einstein tensor, $\Theta_{\xi\eta}= g^{\mu\nu}\frac{\delta
\mathrm{\textit{T}}_{\mu\nu}}{\delta g_{\xi\eta}}$, $f\equiv f(\mathcal{G},T)$, $f_{\mathcal{G}}\equiv\frac{\partial f (\mathcal{G},\mathrm{\textit{T}})}{\partial
\mathcal{G}}$, and $f_{\mathrm{\textit{T}}}\equiv\frac{\partial
f(\mathcal{G},\mathrm{\textit{T}})}{\partial \mathrm{\textit{T}}}$.
  It is worth mentioning here that the Einstein equations can be resuscitated by simply putting
$f(\mathcal{G},\mathrm{\textit{T}})=0$ whereas the field equations for
$f(\mathcal{G})$ can be revived by replacing
$f(\mathcal{G},\mathrm{\textit{T}})$ with $f(\mathcal{G})$ in Eq.($\ref{4_eqn}$).
The expression for energy-momentum tensor $\mathrm{\textit{T}}_{\xi\eta}^{(m)}$ is as follows
 \begin{equation}\label{emt}
\mathrm{\textit{T}}_{\xi\eta}^{(m)}=-\frac{2}{\sqrt{-g}}\frac{\delta(\sqrt{-g}\mathcal{L}_{M})}{\delta
g^{\xi\eta}}.
\end{equation}

However the energy-momentum tensor depending on the metric has the form
\begin{equation}\label{emt1}
\mathrm{\textit{T}}_{\xi\eta}^{(m)}=g_{\xi\eta}\mathcal{L}_{M}-2\frac{\partial\mathcal{L}_{M}}{\partial
g^{\xi\eta}}.
\end{equation}\par
The description of the covariant divergence of Eq.(\ref{4_eqn}) reads
\begin{equation}\label{div}
\nabla^{\xi}T_{\xi\eta}=\frac{f_{\mathrm{\textit{T}}}(\mathcal{G},\mathrm{\textit{T}})}
{\kappa^{2}-f_{\mathrm{\textit{T}}}(\mathcal{G},\mathrm{\textit{T}})}\bigg[(\mathrm{\textit{T}}_{\xi\eta}+\Theta_{\xi\eta})
\nabla^{\xi}(\text{ln}f_{\mathrm{\textit{T}}}(\mathcal{G},\mathrm{\textit{T}}))+
\nabla^{\xi}\Theta_{\xi\eta}-
\frac{g_{\xi\eta}}{2}\nabla^{\xi}T\bigg].
\end{equation}
The effect of divergences on this modified theory of gravity may not be avoided as it has already happened to all the other modified theories of gravity. These divergences come into existence because of the natural involvement of the higher order derivatives terms of the stress-energy tensor. In fact, this intricate situation seems to be unavoidable, but the weak equivalence norms must remain uncompromising with such modifications to the Einstein theory by the addition of the auxiliary fields. However, the new constraints to the Eq.(\ref{div}) may be considered to acquire the standardized expression for the matter stress-energy tensor.\par

Given below is the energy-momentum tensor $T_{\xi\eta}$ defined for the perfect matter source as
\begin{equation}\label{12}
T_{\alpha\beta}=(\rho+p)\xi_{\alpha}\xi_{\beta}-{p}g_{\alpha\beta},
\end{equation}
where $p$ and $\rho$ are the pressure and energy density, respectively.
The 4-velocity vector denoted by $\xi_{\zeta}$, satisfies

\begin{equation}\label{13}
\xi^{\alpha}\xi_{\alpha}=1, ~~~and~~~\xi^{\gamma}=e^{\frac{-\varphi}{2}}\delta^{\gamma}_{1}.
\end{equation}
It may be noted that we have assumed the substitute of lagrangian matter as $\mathcal{L}_{M}=-p$, therefore, $\Theta_{\zeta\eta}$ turns out to be

\begin{equation}\label{Bigtheta}
\Theta_{\alpha\beta}=-2T_{\alpha\beta}-p g_{\alpha\beta}.
\end{equation}
For this particular investigation of the stellar hydrostatic equilibrium compact structures, we opt for $f(\mathcal{G},T)$ gravity model, given as
\begin{equation}\label{VM}
f(\mathcal{G},T)=f_{1}(\mathcal{G})+f_{2}(T),
\end{equation}
where $f_{1}(\mathcal{G})$ is a single variable function of the Gauss-Bonnet term $\mathcal{G}$ with $f_{1}(\mathcal{G})=\alpha\mathcal{G}^n$, a power law $f(\mathcal{G})$ model as proposed by Cognola et al. \cite{17}, with $\alpha$ being an any real constant, and $n$ a positive real number.  Here for the second part, we set $f_{2}({T})=\lambda\textit{T}$, with $\lambda$ a real coupling constant, the role of which has to be very critical in distinguishing the outcomes related to the compact stars to those with the results in GR. We take here $\alpha=1$ and $n=1$ for our upcoming study. For this model under consideration, the field equations (\ref{4_eqn}) read
\begin{equation}\label{modelEQ}
G_{\zeta\eta}=(\lambda+1)T_{\zeta\eta}+\frac{1}{2}g_{\zeta\eta}[\lambda(2p+T)+\mathcal{G}].
\end{equation}
Just to mention here that the general field equation in GR can be retrieved by simply substituting $\alpha=\lambda=0$. Now under the new circumstances, Eq.(\ref{div}) takes the shape as
\begin{equation}\label{modiv}
\nabla^{\zeta}T_{\zeta\eta}=\frac{\lambda}{\lambda+1}\bigg[\nabla^{\zeta}(pg_{\zeta\eta})+\frac{1}{2}g_{\zeta\eta}\nabla^{\zeta}T\bigg].
\end{equation}
It is quite simple to observe that by taking $\lambda=0$ in Eq.(\ref{modiv}),  the original conserved form in GR of the energy-momentum tensor is obtained.\par

\section{Equations of Stellar Configurations in $f(\mathcal{G},T)$ Gravity}
The densest cold objects exhibit a huge gravitational collapse because their immense mass surpasses the critical limit. It is too hard to halt this ultimate collapse for the matter with the existing high (EoS) parameter. Next, we find out some important stellar expressions to investigate the physical aspects of the stellar $f(\mathcal{G},T)$ gravity model.
\subsection{Stellar Equilibrium Equation}

For the purpose to specifically develop the $f(\mathcal{G},T)$ hydrostatic compact equilibrium equation, we take into account the uncharged, non-rotating spherically symmetric metric as

\begin{equation}\label{11}
ds^{2}=e^{\psi(r)}dt^{2}-e^{-\varphi(r)}dr^2-r^{2}(d\theta^{2}+sin^{2}\theta d\phi^{2}),
\end{equation}
where $\psi$ and $\varphi$ being some arbitrary function of radius $r$.

For the metric (\ref{11}), the non-null components of Einstein tensor $G_{\zeta\eta}$ read
\begin{equation}\label{G00}
  G_{00}=\frac{e^{\psi -\varphi } \left(r \varphi '+e^{\varphi }-1\right)}{r^2},
\end{equation}
\begin{equation}\label{G11}
G_{11}= \frac{r \psi '-e^{\varphi }+1}{r^2},
\end{equation}
\begin{equation}\label{G22}
G_{22}=\frac{1}{4} r e^{-\varphi } \left(-r \varphi ' \psi '+2 r \psi ''+r \left(\psi '\right)^2-2 \varphi '+2 \psi '\right).
\end{equation}
where $'$  denotes the radial derivative. Now manipulating Eqs.(\ref{12}), (\ref{G00}-\ref{G22}), and (\ref{modelEQ}), we get

\begin{eqnarray}\nonumber
&&2+r^{2}(p\lambda-(2+3\lambda)\rho)=\frac{e^{-2\varphi}}{4}[r^{2}\psi'^{4}-2r^{2}\psi'^{3}\varphi'+8e^{\varphi}(-1+r\varphi'+4\psi'')+\\\nonumber
&&4\psi''(-8+r^{2}\psi'')-4\psi'\varphi'(4(-3+e^{\varphi})+r^{2}\psi'')+
\psi'^{2}(-8+16e^{\varphi}+r^{2}\varphi'^{2}+4\psi'')]\\\label{G0}\\\nonumber,
&&2+e^{\varphi}(-2+r^{2}(-p(2+3\lambda)+\lambda\varphi))=\frac{-e^{-\varphi}}{4}[-r^{2}\psi'^{4}+2r^{2}\psi'^{3}\varphi'-8\varphi'^{2}-\\\nonumber
&&4\varphi''(-8+8e^{\varphi}+r^{2}\psi'')-\psi'^{2}(16(-1+e^\varphi)+r^{2}(\varphi'^{2}+4\psi''))+4\psi'(2e^{\varphi}r+\\\label{G1}
&&\varphi'(4(-3+e^{\phi})+r^{2}\psi''))],\\\nonumber
&&8+e^{\varphi}(-4+r^{4}(-p(2+3\lambda)+\lambda\rho))=\frac{-e^{-\varphi}}{2}[8+r^{2}(6\psi'^{2}-24\psi'\varphi'-\\\label{G2}
&&2\varphi'^{2}+16\psi''+e^\varphi ((2r+(-8+r^{2})\psi')(\psi'-\varphi')+2(-8+r^{2})\psi''))].
\end{eqnarray}

Now for the $f(\mathcal{G},T)$ gravity, the non-conservation Eq.(\ref{div}) of energy-momentum tensor takes the new shape as
\begin{eqnarray}\label{Ndiv}
0=\frac{dp}{dr}+\frac{\psi'}{2}(\rho+p)+\frac{\lambda}{\lambda+1}(p'-\rho').
\end{eqnarray}
The investigations pertaining to the static and spherically symmetric compact objects are carried out with the assumption of the exterior spacetime solution of the star is defined by a Schwarzschild’s metric. In GR, the incredible predictions of Schwarzschild’s geometry have continued to be pioneer in choosing from the diverse matching possibilities while exploring stellar compact stars exterior solutions. Now in turn to the modified theories of gravity, the modified Tolman-Oppenheimer-Volkoff (TOV) equations \cite{OV} with negligible pressure and energy density, the outer solution of the star may deviate from the Schwarzschild's solution. Nevertheless, it is anticipated that the modified TOV solutions with pressure and energy density (may be non-zero) may acknowledge Schwarzschild's solutions with some appropriate choice of $f(\mathcal{G},T)$ gravity compact stellar model. Maybe, due to this assumption, Birkhoff’s theorem may not be accepted in modified gravity. The comprehensive analysis of this phenomenon and to understand the interconnected concerns in context of modified $f(\mathcal{G},T)$ gravity might prove a captivating pursuit. Many authors followed the Schwarzschild exterior solutions to undergo such explorations and established some interesting outcomes \cite{27a}-\cite{Ast}.\\\
Thus, it appears fairly motivating here to derive TOV equations for $f(\mathcal{G},T)$ stellar model \cite{Sir&Me5}. For this, we take the radial gravitational mass $m(r)$ of a sphere with inner radii $r$ such that $e^{-\varphi}=1-\frac{2m}{r}$. Now working with Eqs.(\ref{G0}) and (\ref{G1}) together with the gravitational mass, one reads

\begin{equation}\label{AD}
  \psi'=-\frac{\frac{2 m^2}{r-2 m}+r m'}{r^4}-\frac{(\lambda +1) r^2 (p+\rho )}{2 m-r},
\end{equation}
Manipulating Eqs.(\ref{G0}-\ref{G2}) together with $e^{-\varphi}=1-\frac{2m}{r}$, the differentiation of the gravitational mass with respect to the radius $r$, gives
\begin{equation}\label{Gmass}
 \frac{dm}{dr}= \frac{2 m \left(m+r^3\right)-(\lambda +1) r^6 (p+\rho )}{r \left(2 m+2 r^3-r\right)}.
\end{equation}
Solving Eq.(\ref{AD}) together with Eq. (\ref{Ndiv}), one finds the hydrostatic equilibrium equation in $f(\mathcal{G},T)$ gravity, as

\begin{eqnarray}\label{mtov}
\frac{dp}{dr}=\frac{-1}{2}(\rho+p)\frac{\frac{m}{r^{4}}-r^{2}\frac{(1+\lambda)(p+\rho)}{(2m-r)}-\frac{\frac{dm}{dr}}{r^{3}}}{(1-\frac{d\rho}{dp})(1+\frac{\lambda}{1-\lambda})}.
\end{eqnarray}

We have imposed here the condition that the energy density $\rho$ is dependent on the pressure $p$ such that $\rho=\rho(p)$. It is noted here that by taking $\lambda=0$, one can work out the TOV equation related to the case of GR.\par
\subsection{The Boundary Conditions}
We are certainly interested here to integrate the differential Eqs.(\ref{Gmass}) and (\ref{mtov})together for the purpose to explore some important physical characteristics of the hydrostatic stellar equilibrium of the compact stars. The ultimate solution of these equations may not be presented  until we define some specific boundary conditions starting with the center of the star at $r=0$ \cite{MOR}:
\begin{equation}\label{BCs}
  m(0)=0,~~~\rho(0)=\rho_{c},~~~p(0)=p_{c}.
\end{equation}
The solutions at the surface of the compact stars ($r=R$) are evaluated with the condition of $p(R)=0$ in such a way that the interior spacetime of the star is matched smoothly to the exterior Schwarzschild spacetime. The interior and the exterior spacetime potential metrics are connected through $e^{\psi(R)}=e^{-\varphi(R)}=1-\frac{2M}{R}$, with $M$ as the total stellar mass of the star.
 \subsection{Equation of State Models}
 The structure and the formation the compact stars particularly of a neutron star is totally dependent on the equation of state (EOS) parameter which establishes the relation between the pressure and the energy density in the interior of the star \cite{Latt}. Till to date, the physical attributes of the stiff matter under these specific extreme circumstances can only be premeditated on the basis of physical theoretical models. So far, to conceive the required exceptionally high energy density in the accelerator experiments has not been possible. However, with the consideration of an EOS, a stellar mass-radius relationship for the compact stars corresponding to their maximum mass may be derived.\\\
  Therefore, now for the purpose of making the system of equations closed, the appropriate choice of the EoS becomes significant here. To carry on with some simplest choice of the polytropic EoS, the work by Tooper \cite{PRD22} may be a good example and it can be followed by considering $p=\omega\rho^{5/3}$, with $\omega$ being the EoS parametric constant. Now using this relation together with the equations (\ref{Gmass}) and (\ref{mtov}), one reads
  \begin{eqnarray}\label{EQ1}
0&=&\frac{(\rho+\rho^{5/3}\omega)}{2(1+\frac{\lambda}{1-\lambda})(1-\frac{3}{5\rho^{2/3}\omega})}\bigg(-\frac{r^2\rho(1+\lambda)(1+\rho^{2/3}\omega)}{2m-r}\\\nonumber
&-&\frac{2m(m+r^3)-r^6(1+\lambda)\big(\rho+\rho^{5/3}\omega\big)}{r^4(2m-r+2r^3)}+\frac{m}{r^4}\bigg)+\frac{5}{3}\rho^{2/3}\omega \frac{d\rho}{dr}.
\end{eqnarray}
    As in \cite{PRD23,PRD24}, one may take $\omega=1.4745\times10^{-3}[fm^3/MeV]^{2/3}$.  Like the analysis of compact stars in $f(R,T)$ theory of gravity, the discussion on their formation, internal and external structures, composition and evolution can be done in $f(\mathcal{G}, T)$ theory of gravity. This can be accomplished by investigating the hydrostatic equilibrium configurations of the stars.\par
    When the strange quark matter (SQM) is under examination, then the choice of the MIT bag model has to be more appropriate. This is due to the fact that chosen EoS exhibits the composition of fluid related to strange quarks only [42]. It is described by the expression $p=\alpha(\rho-4\beta)$ and is used for the investigations of the stellar configurations of the compact objects \cite{{Farhi},{Malheiro2}} with $\alpha$, and $\beta$ being the constants. Now for this situation, equations (\ref{Gmass}) and (\ref{mtov}) constitue
    \begin{eqnarray}\label{EQ2}
0&=&\frac{d\rho}{dr}+\frac{(\alpha(\rho-4\beta)+\rho)}{2\big(\frac{\lambda}{1-\lambda}+1\big)\big(1-\frac{1}{\alpha}\big)}\bigg(\frac{m}{r^4}-\frac{(\lambda+1)r^2(\alpha(\rho-4\beta)+\rho)}{2m-r}\\\nonumber
&-&\frac{2m(m+r^3)-(\lambda+1)r^6(\alpha(\rho-4\beta)+\rho)}{r^4(2m+2r^3-r)}\bigg).
\end{eqnarray}
       The value of $\alpha$ for the massive quarks with $m_{S}\approx250[MeVfm^{-3}]$ equals to 0.28 and for the strange quarks without mass, it is taken as $1/3$. \par

   After we have defined our preference for the EoS in both the cases, the equations (\ref{EQ1}) and (\ref{EQ2}) are to be dealt separately. One can see that these equations are highly non-linear differential equations and deny any information through analytic solutions. Therefore, one may opt for some suitable numerical methods depending on their computational efficiency for this specific situation. We have preferred the Runge-Kutta $4$th order method to determine three unknown $m$, $\rho$, and $p$, all being the radial functions.\par
   The investigations on the compact stars, proton stars in particular, reveal that they consist of some densest structures of matter in this existing cosmological universe.  Compactness of these objects depends on some constraints which are set by GR and the causality limit. Moreover, the analytic solutions provided by GR are no doubt very handy in comprehending to some extent the relations among the maximal mass, energy density, radial pressure, radii, and moments of inertia of the compact stars. But there is still uncertainty about the non-dependency of some of these relations with the dense matter equation of the state while the others being dependent on the equation of the state. There are certain constraints on the structures of the neutron stars in connection with the equation of state and are set by recent interpretations from the sources like pulsar timing, Type I X-ray bursts, and from binaries X-ray emissions. Similar conditions suggest the existence  of some strange quark matter stars in the cores of some typical neutron stars.
\section{Solutions of Neutron and Strange Stars in $f(\mathcal{G},T)$ Gravity}
We have just settled things to work out the neutron and the strange star solutions numerically by using the Runge-Kutta method. This will be done along with the boundary conditions for some different values of the central density $\rho_{c}$ and the parametric values of $\lambda$ in $f(\mathcal{G},T)$ gravity model. It is important to mention here that for $\lambda=0$, the results will correspond to those in GR and for the non-zero parametric values of $\lambda$, they will relate to the modified $f(\mathcal{G},T)$ theory of gravity. Now under these circumstances we discuss now the different physical aspects of the compact stars like the energy density, pressure and the stellar mass in different situations.
\subsection{Energy Density and Pressure Profile}
Using $f(\mathcal{G},T)$ gravity stellar model, we have investigated three important physical constituents of the compacts stars i.e. energy density $\rho$, the pressure $p$, and the normalized stellar mass $m/M_{\odot}$ as a radial function. Two different EoS have been used with their corresponding representation to the neutron stars and the strange stars as it is expressed in two different equations (\ref{EQ1}), and (\ref{EQ2}), respectively. In this work, the left profiles are for the neutron stars and right ones are for the strange stars using different values of $\lambda$ and the value of the central energy density  as $900[MeV(fm^{-3})]$.\par
\begin{figure}\center
\fontsize{06}{08}\selectfont
\begin{tabular}{cccc}
\\ &
\epsfig{file=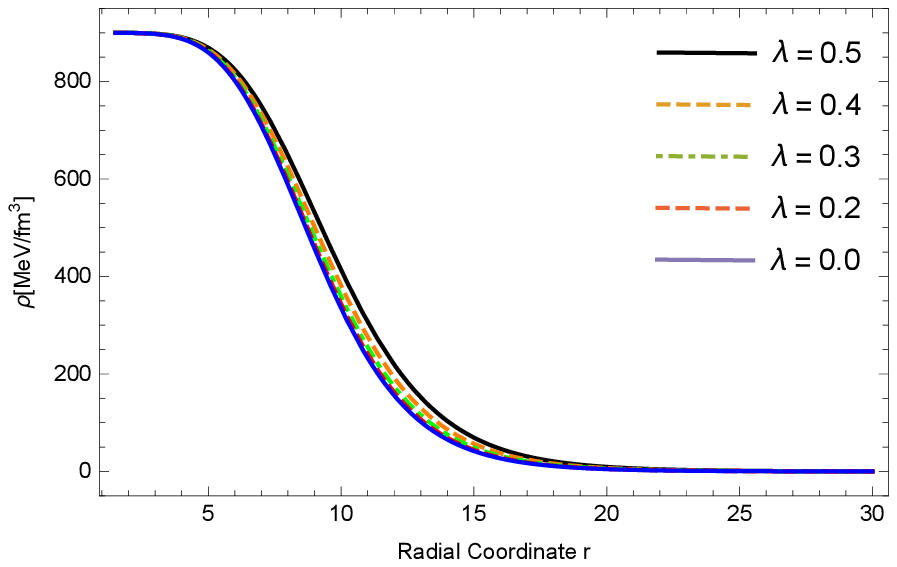,width=0.50\linewidth} &
\epsfig{file=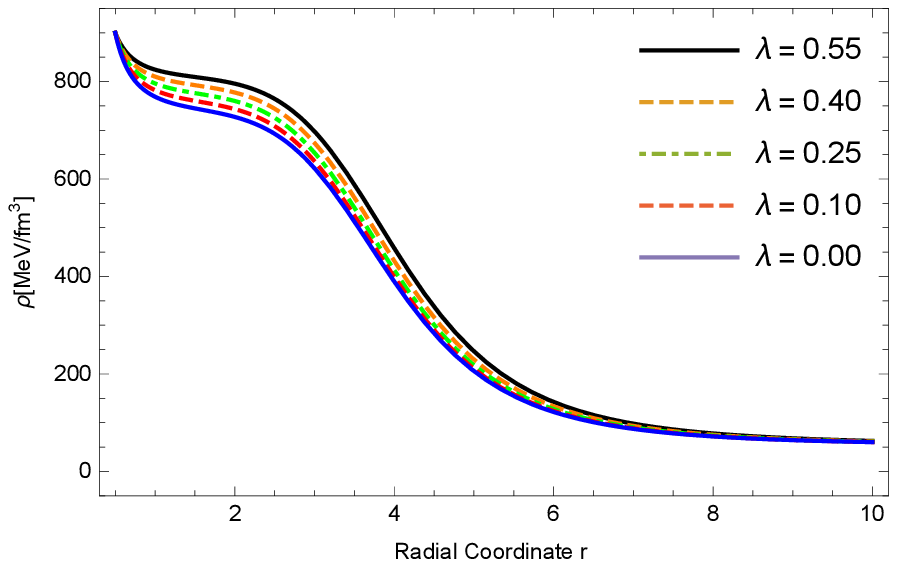,width=0.50\linewidth}\\
\end{tabular}
\caption{The energy density profiles are presented with respect to the radial coordinate $r$ ($km$) for the neutron stars on the left panel and for the strange stars on the right panel, for some different parametric values of $\lambda$. Here, the value the cental energy density has been taken as $900[MeV/fm^3]$.\label{fig:drpf}}\center
\end{figure}
\begin{figure}\center
\begin{tabular}{cccc}
\\ &
\epsfig{file=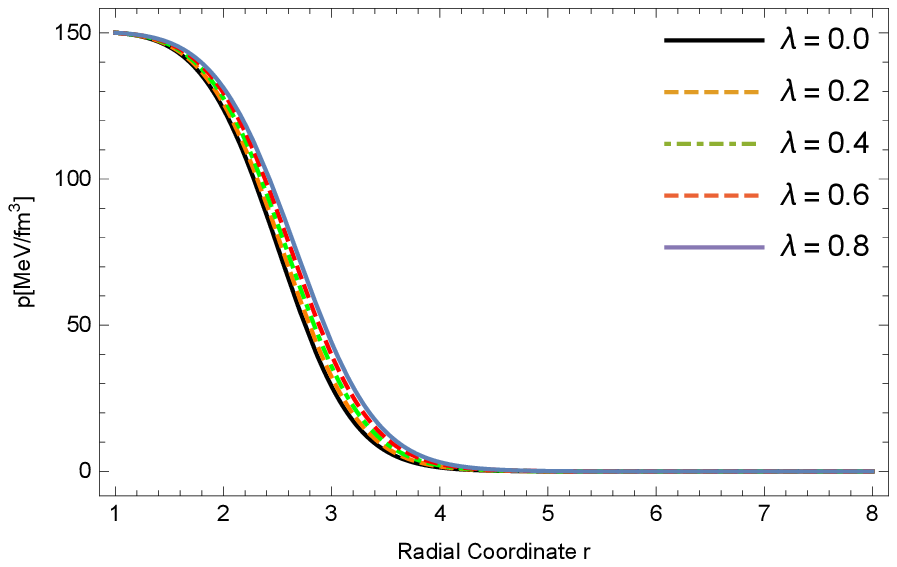,width=0.5\linewidth} &
\epsfig{file=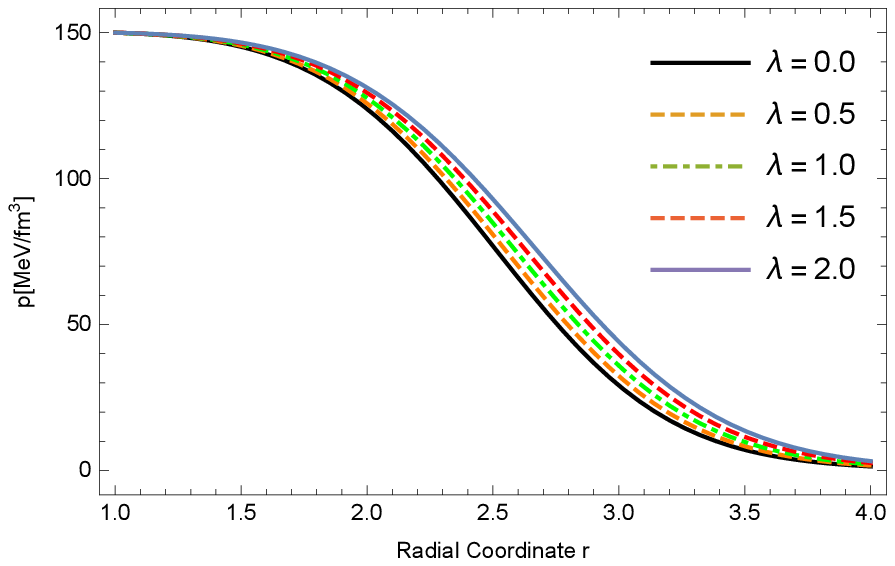,width=0.5\linewidth}\\
\end{tabular}
\caption{The pressure profiles are presented with respect to the radial coordinate $r$ ($km$) for the neutron stars on the left panel and for the strange stars on the right panel, for some different parametric values of $\lambda$. Here, the value the cental energy density has been taken as $900[MeV/fm^3]$.\label{fig:prpf}}\center
\end{figure}
\begin{figure}\center
\begin{tabular}{cccc}
\\ &
\epsfig{file=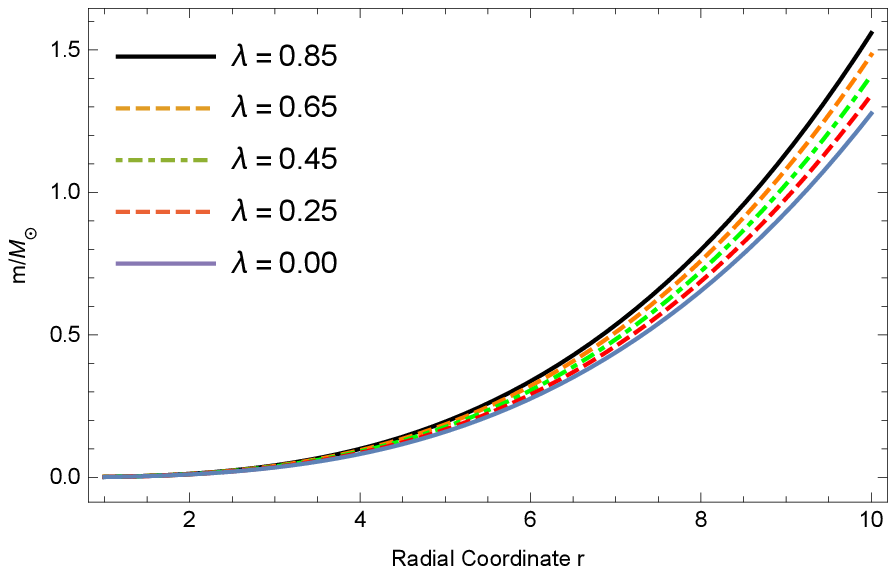,width=0.5\linewidth} &
\epsfig{file=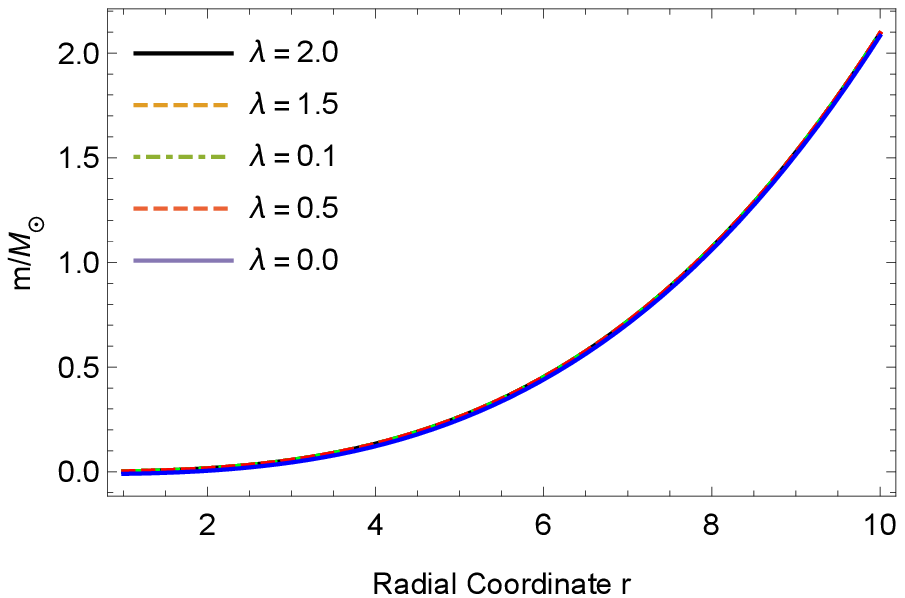,width=0.5\linewidth}\\
\end{tabular}
\caption{The profiles of the masses normalized with the solar masses are presented with respect to the radial coordinate $r$ ($km$) for the neutron stars on the left panel and for the strange stars on the right panel, for some different parametric values of $\lambda$. Here, the value the cental energy density has been taken as $900[MeV/fm^3]$.\label{fig:mrpf}}\center
\end{figure}

The behavior of $\rho$, $p$, and $m/M_{\odot}$ is quite evident from the Figures (\ref{fig:drpf}-\ref{fig:mrpf}) when manipulated with some different values of $\lambda$. One can observe from these plots that as $r\rightarrow0$, $\rho$ goes to maximum, and this in fact shows the high compact nature of the core of the star. Both of the plots in Figure (\ref{fig:prpf}) indicate that the radial pressure for neutron stars and the strange stars is  decreasing  with the increasing radius $r$, which further confirms the compact nature of the stars. Moreover, the character of the term $\lambda{T}$ plays a significant role here when one observes the increasing stellar mass of the compacts stars with the positively increasing $\lambda$ term. This situation may come up with a comment that this increasing effect caused by the $\lambda$ parameter is very similar to the extra electric charge or pressure for the neutron and strange star structures in GR \cite{SRAY}-\cite{Negreiro}. Additionally, it is worth mentioning here that the mass-radius relation depicts a directly proportional relation which is physically attributed to the compact objects, as well. In consideration of the foregoing, the total stellar radius of the neutron stars keeps on growing with the increasing $\lambda$ values, meanwhile, the total radius of the strange stars slowly keeps on decreasing with the increasing values of $\lambda$. This opposite kind of behavior of the stellar radius of the neutron stars and strange stars with the increasing $\lambda$ requires some more explanation, to be given in next subsection.

\subsection{Equilibrium Configurations of the Compact stars}
\begin{figure}\center
\begin{tabular}{cccc}
\\ &
\epsfig{file=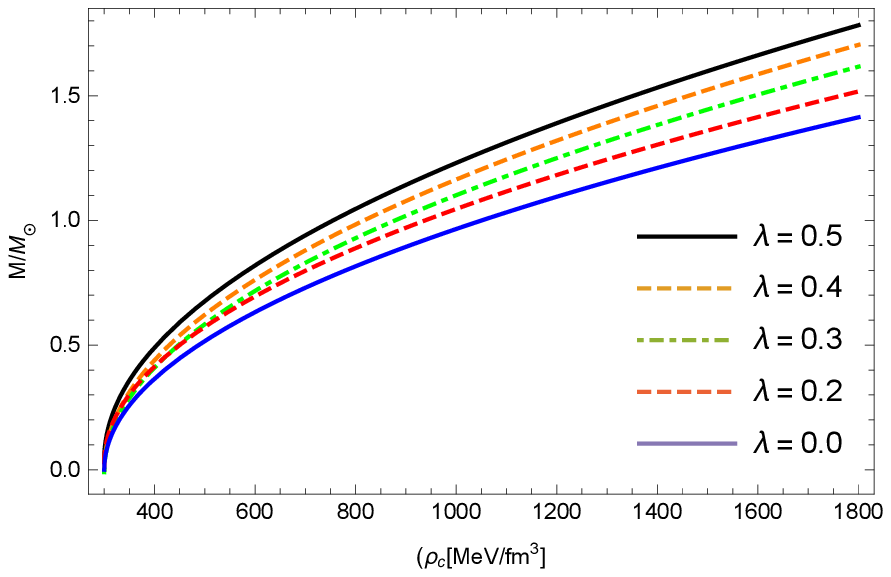,width=0.5\linewidth} &
\epsfig{file=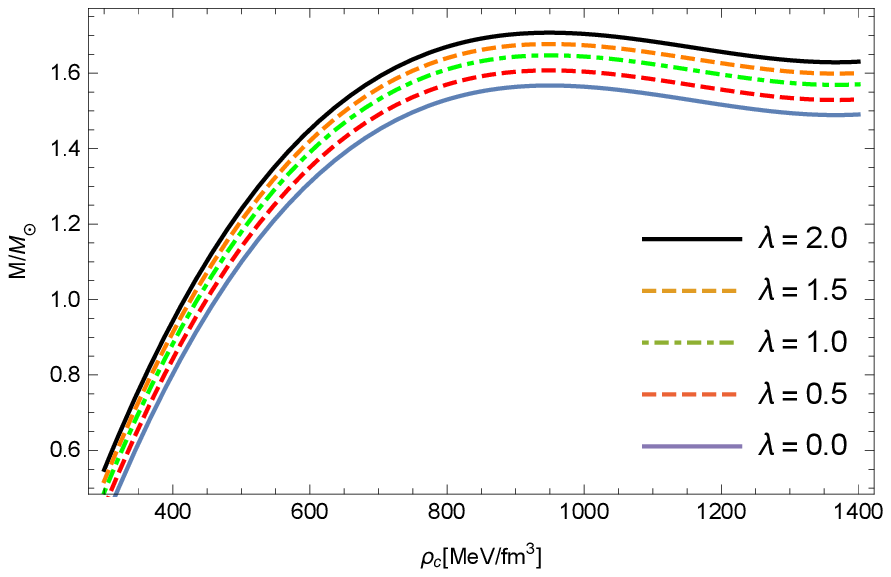,width=0.5\linewidth}\\
\end{tabular}
\caption{The profiles of the total stellar normalized masses($M/M_{\odot}$) depending on the central energy density $[MeV/fm^3]$ are presented for the neutron stars on the left panel and for the strange stars on the right panel, for some different parametric values of $\lambda$.\label{fig:Mdpf}}\center
\end{figure}

The profiles of the total stellar masses normalized with the solar masses both for the neutron stars (left plot) and for the strange stars (right plot) are presented in Figure (\ref{fig:Mdpf}) with respect to the central density $\rho_{c}$  for some different parametric values of $\lambda$. The masses for both of the stars show an increasing and steady positive behavior with the increasing $\rho_{c}$. This increasing behavior tends to the point where the total stellar masses reach their maximum value until they start decreasing monotonically with the increasing $\rho_{c}$. Now the role of the coupling parameter $\lambda$ is interesting here as well, as it can be observed that if the positive increments in its values are kept on increasing, the stellar masses go on increasing too. Therefore, the larger stellar masses are obtained for the larger values $\lambda$, but with comparatively smaller central densities. Also, a maximum stellar mass point against a lower central energy density is obtained. \par
For the neutron stars, when $\lambda=0$, the maximum stellar mass reaches $1.489M_{\odot}$ against the value of central density  $\rho_{c}\thickapprox16.296~\rho_{nuclear}$. With the increasing values of $\lambda$, we noted the corresponding increase in the maximum stellar masses. For $\lambda=0.5$, the maximum stellar mass value $1.98M_{\odot}$ was determined when $\rho_{c}\thickapprox16.563~\rho_{nuclear}$. On the other hand, for the case of the strange stars, depicted in the right plot shows similar behavior but obviously with numerically different values. When we substitute $\lambda=0$, then the maximum stellar mass obtained was $1.41M_{\odot}$ against $\rho_{c}\thickapprox6.691~\rho_{nuclear}$. When $\lambda=0.5$, and for $\rho_{c}\thickapprox6.537~\rho_{nuclear}$, the maximum mass value reached $1.85M_{\odot}$.\par
An important understanding which ultimately develops during this discussion investigating the neutron stars is the substantial dependance of the stellar mass on the EoS for some particular choice of $\lambda$, and this may be extended for the better understanding of pulsars PSR J$0348+0432$ and PSR J$1614-2230$ which are expected to have the higher masses. Such discussions can be noted in the case of $f(R,T)$ gravity in \cite{MOR}.\par
\begin{figure}\center
\begin{tabular}{cccc}
\\ &
\epsfig{file=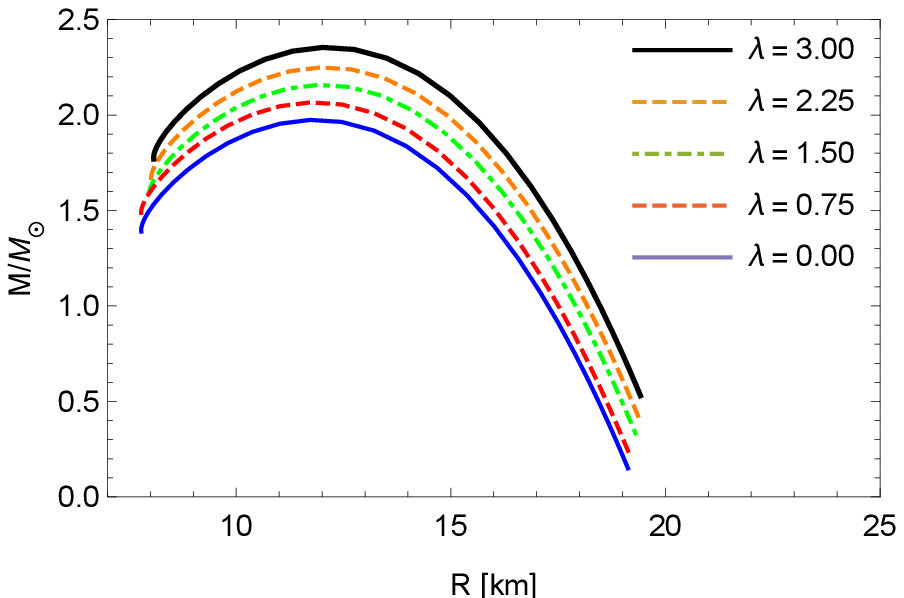,width=0.5\linewidth} &
\epsfig{file=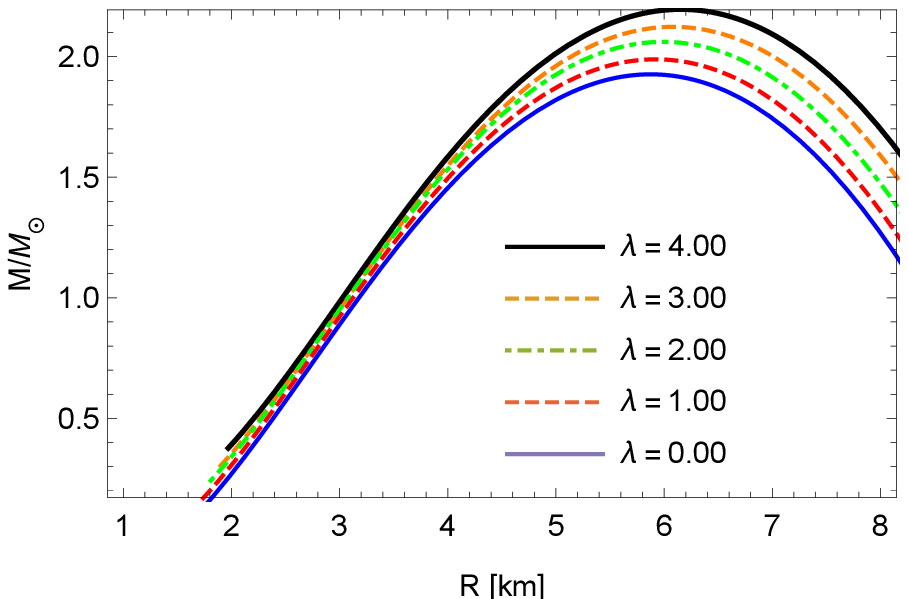,width=0.5\linewidth}\\
\end{tabular}
\caption{The profiles of the total stellar normalized masses($M/M_{\odot}$) as a function of the total radius $R$ ($Km$) are presented for the neutron stars on the left panel and for the strange stars on the right panel, for some different parametric values of $\lambda$.\label{fig:NMRf}}\center
\end{figure}

In Figure \ref{fig:NMRf} profiles of the total stellar normalized masses($M/M_{\odot}$) depending on the total radius $R$ ($km$) are shown for the neutron stars in the left plot and for the strange stars on the right plot attributed for two different choices of the EoS and for some distinct parametric values of $\lambda$. It can be seen that with the increasing values of $\lambda$, the neutron stars get bigger in size and massive. Similar corresponding behavior is noted for the strange stars.
\begin{figure}\center
\begin{tabular}{cccc}
\\ &
\epsfig{file=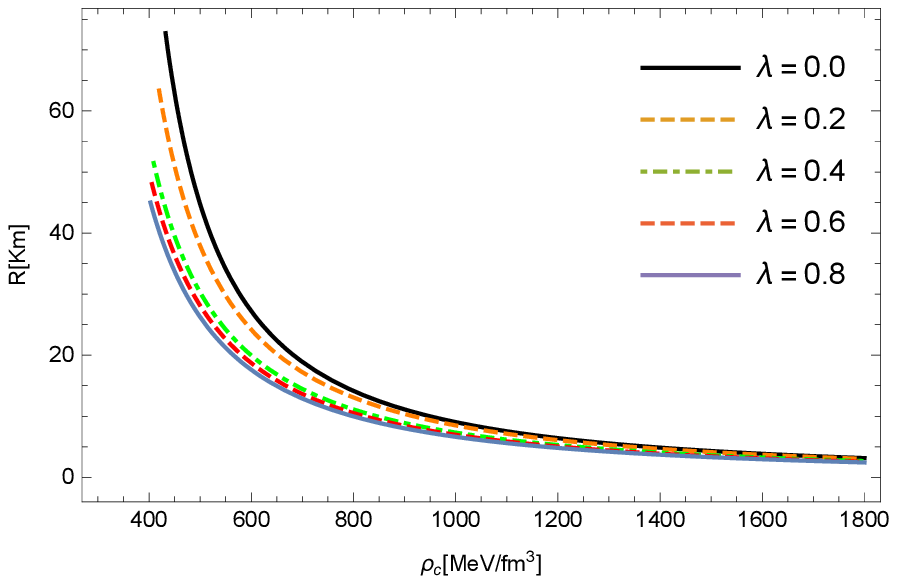,width=0.5\linewidth} &
\epsfig{file=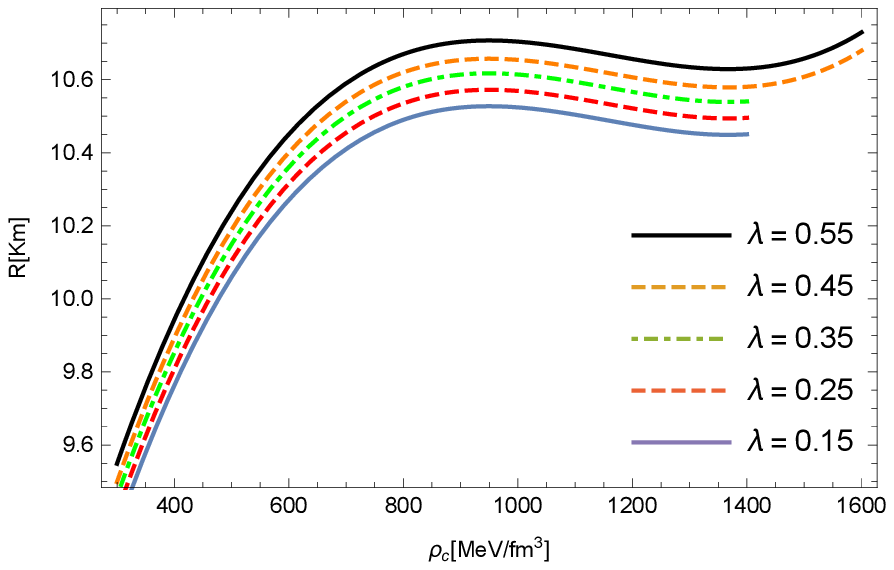,width=0.5\linewidth}\\
\end{tabular}
\caption{ The profiles of the total stellar radius $R$ ($Km$) depending on the central energy density $[MeV/fm^3]$ are presented for the neutron stars on the left panel and for the strange stars on the right panel, for some specific parametric values of $\lambda$.\label{fig:Rdf}}\center
\end{figure}
\subsection{Total Stellar Radius of Compact Stars}

In Figure (\ref{fig:Rdf}) the profiles of the total stellar radius $R$ ($Km$) as a function of central energy density $\rho_{c}$ $[MeV/fm^3]$ are shown for the neutron stars on the left plot and for the strange stars on the right plot for the choice of some different parametric values of $\lambda$. It is quite interesting to note here that for the case of the neutron stars the total stellar radius increases while for the case of the strange stars it decreases, with the increasing values of the coupling parameter $\lambda$. This opposite behavior is apparently due to the lessening effect of the negative radial derivative of the pressure in neutron stars as $\lambda$ keeps on increasing. Therefore, it can be further concluded as the radius of the star gradually increases, the pressure gets slower and slower till the bigger stellar total radius is obtained. Contrary to this, for the case of the strange stars, as the radial coordinate keeps on increasing with the increasing $\lambda$, the negative $d\rho/dr$ gets larger resulting into the faster decay of the pressure term. Hence a lower total radius is achieved.

\section{Concluding Discussion}
Compact stars originate from the final phase of the gravitational collapse during the evolution process of some gigantic ordinary stars due to their failure to sustain their stability as they would have, at that time, ceased their nuclear fuel important for their existence. This is the occasion when the internal pressure no longer is sufficient to overcome the gravitational force and this causes the star to collapse leaving behind compact objects like the neutron stars, strange stars, white dwarfs, or a black hole. These compact stars are made of some matter with high densities responsible for their massive structures. This is what which makes these compact stars, neutron stars in particular, highly massive objects with very small radii. These structures are geometrically described by the TOV equations, the solution of which becomes too important to study the complexities involved in these massive small-sized structures. These equations constitute the mass, radius, energy density, and pressure and show how the pressure and the energy density variate with respect to the mass and radius of the compact star. In case of compact stars (in particular the neutron stars), the internal pressure equivalent to the gravitational pressure is in fact the pressure being exhibited from some degeneracy of fermions. The desired solution of TOV equations totally depend on some appropriate choice of the equation of state which actually tells us the proportional relation between the energy density and the pressure.\par
Despite the huge success of GR, the role of modified theories of gravity in investigating the dynamics of the gravitational collapse and instability range have become important. Theories of gravity like $f(R,T)$, $f(R)$, $f(G)$ have contributed in understanding the structures of the compact stars and matter with high densities. There are growing expectations from these modified theories to work out some self-consistent compact structure model obtained from some exact solution of TOV equations, and be pragmatic with some specific EoS.\par
Here, in this paper, we have investigated the hydrostatic equilibrium static configurations of the neutron stars and the strange stars under $f(\mathcal{G},T)$ gravity by developing a TOV equation for the theory. For this purpose, we have taken into account  EoS models, the polytropic model, $p=\omega{\rho^{5/3}}$ and the MIT bag model, $p=0.28(\rho-4\beta)$ for the neutron stars and the strange stars, respectively. The presence of the term $\lambda$ is due to the addition of an extra term $\lambda{T}$ in the functional form of $f(\mathcal{G},T)$ which plays a significant role in this analysis.\par
For our $f(\mathcal{G},T)$ gravity model, depending on the different values of $\lambda$, the behavior of $\rho$, $p$, and $m/M_{\odot}$ has been studied and is depicted in Figures (\ref{fig:drpf}-\ref{fig:mrpf}). It is evident from these plots both for the neutron stars (left) and for the strange stars (right) that as $r\rightarrow0$, $\rho$ goes to maximum showing the high compact nature of the core of the star, while the radial pressure for both of the stars decreases with the increasing radius $r$ as shown in Figure (\ref{fig:prpf}) confirming again this compact nature, with the increasing $\lambda$.\par

We have observed that for the case of neutron stars, when $\lambda=0$, the maximum stellar mass $1.489M_{\odot}$ is obtained when the central density reaches $\rho_{c}\thickapprox16.296~\rho_{nuclear}$. Moreover, for the increasing $\lambda$, we found the relative increase in the maximum stellar masses. When $\lambda=0.5$, the maximum stellar mass point $1.98M_{\odot}$ was achieved with the reading of $\rho_{c}\thickapprox16.563~\rho_{nuclear}$. In turn, for the strange stars case as depicted in the right plot of Figure (\ref{fig:Mdpf}) shows similar behavior but with numerically different values. When we substitute $\lambda=0$, then the maximum stellar reached $1.41M_{\odot}$ against $\rho_{c}\thickapprox6.691~\rho_{nuclear}$. When $\lambda=0.5$, and for $\rho_{c}\thickapprox6.537~\rho_{nuclear}$, the maximum mass value reached $1.85M_{\odot}$.\par
An important understanding which ultimately develops during this discussion investigating the neutron stars is the substantial dependance of the stellar mass on the EoS for some particular choice of $\lambda$, and this may be extended for the better understanding of pulsars PSR J$0348+0432$ and PSR J$1614-2230$ which are expected to have the higher masses.\par

While investigating the total stellar radius for the case of the neutron stars, we observed that it increases while in case of the strange stars it decreases, with the increasing  $\lambda$. This different behavior is in fact because of the lessening effect of the negative radial derivative of the pressure in neutron stars as $\lambda$ keeps on increasing. Therefore, it can be further concluded as the radius of the star gradually increases, the pressure gets slower and slower till the bigger stellar total radius is obtained. Contrary to this, for the case of the strange stars, as the radial coordinate keeps on increasing with the increasing $\lambda$, the negative $d\rho/dr$ gets bigger causing a faster decay of the pressure term. Hence a lower total radius is acquired.\par


\end{document}